\documentstyle[12pt,epsf]{article}
\begin{document}

\begin{flushright}
USM-TH-97
\\
 CPT-2000/P.4045\\
\end{flushright}
\bigskip\bigskip

\centerline{\large \bf  Helicity and Transversity Distributions}

\centerline{\large \bf of Nucleon and $\Lambda$-Hyperon from
$\Lambda$ Fragmentation}

\vspace{22pt}
\centerline{\bf
Bo-Qiang Ma\footnote{e-mail: mabq@phy.pku.edu.cn}$^{a}$,
Ivan Schmidt\footnote{e-mail: ischmidt@fis.utfsm.cl}$^{b}$,
Jacques Soffer\footnote{e-mail: Jacques.Soffer@cpt.univ-mrs.fr}$^{c}$,
Jian-Jun Yang\footnote{e-mail: jjyang@fis.utfsm.cl}$^{b,d}$}

\vspace{8pt}

{\centerline {$^{a}$Department of Physics, Peking University,
Beijing 100871, China,}}

{\centerline {CCAST (World Laboratory),
P.O.~Box 8730, Beijing 100080, China,}}

{\centerline {and Institute of Theoretical Physics, Academia
Sinica, Beijing 100080, China}

{\centerline {$^{b}$Departamento de F\'\i sica, Universidad
T\'ecnica Federico Santa Mar\'\i a,}}

{\centerline {Casilla 110-V, 
Valpara\'\i so, Chile}

{\centerline {$^{c}$Centre de Physique Th$\acute{\rm{e}}$orique,
CNRS, Luminy Case 907,}}

{\centerline { F-13288 Marseille Cedex 9, France}}

{\centerline {$^{d}$Department of Physics, Nanjing Normal
University,}}

{\centerline {Nanjing 210097, China}}

\vspace{10pt}
\begin{center} {\large \bf Abstract}

\end{center}
It is shown that $\Lambda$-hyperon fragmentation in charged lepton
deep inelastic scattering (DIS) on a polarized nucleon target can
provide sensitive information concerning the quark helicity and
transversity distributions for both nucleon and $\Lambda$-hyperon
at large $x$. Numerical predictions are given for the spin
transfers of the produced $\Lambda$, when the target nucleon is
polarized either longitudinally or transversely, and with the
nucleon and $\Lambda$ quark distributions evaluated both in an
SU(6) quark-spectator-diquark model and in a perturbative QCD
(pQCD) based model. It is also shown that the predicted spin
transfers have different behaviors for proton and neutron targets,
and this can provide sensitive tests of different predictions for
the quark helicity and transversity distributions of the $d$
valence quark of the proton at large $x$.

\vfill \centerline{PACS numbers: 14.20.-c, 13.85.Ni, 13.87.Fh,
13.88.+e}

\vfill
\centerline{Published in Phys. Rev. {\bf D 64}, 014017 (2001).} 
\vfill

\newpage

Trying to understand the spin content of hadrons is a very
challenging research direction of high energy physics, and so far
many unexpected discoveries have been found in contrast to naive
theoretical considerations. The quark helicity distributions of
the proton $\Delta q(x)$ have been extensively explored in recent
years and our knowledge of them has been considerably enriched.
However, there are still some uncertainties concerning the flavor
decomposition of the quark helicity distributions, especially for
the less dominant $d$ valence quark of the proton. For example,
there are different theoretical predictions for the ratio $\Delta
d(x)/d(x)$ at $x \to 1$: the pQCD based counting rule analysis
\cite{Bro95} predicts $\Delta d(x)/ d(x) \to 1$, whereas the SU(6)
quark-spectator-diquark model \cite{Ma96} predicts $\Delta d(x)/
d(x) \to -1/3$. Available experimental data is not yet accurate
enough to provide a decisive test of the above two different
predictions. On the other hand, our knowledge of the quark
transversity distributions $\delta q(x)$ is very poor, since it is
difficult to measure such quantities experimentally, although
there have been attempts in this direction recently
\cite{HERMES00}. Among some proposals for measuring the quark
transversity distributions, Artru and Mekhfi \cite{Art90}, and
later Jaffe \cite{Jaf96}, have noticed that the $\Lambda$-hyperon
transverse polarization, in the current fragmentation region of
charged lepton deep inelastic scattering (DIS) on the transversely
polarized nucleon target, can provide information of the quark
transversity distribution of the target. However, such a
measurement needs the fragmentation functions of the transversely
polarized quark to transversely polarized $\Lambda$. In the
absence of any theoretical estimate of such quantity, one
possible analysis is to use positivity bounds \cite{DSSV},
but here we will make more specific assumptions.

There has been a suggestion \cite{Lu95} for measuring the nucleon
strange polarizations by the longitudinal $\Lambda$ polarization
in the current fragmentation region of charged lepton DIS on a
longitudinally polarized nucleon target. Such process, as pointed
out by Jaffe \cite{Jaf96}, should be most suitable for extracting
both the quark helicity distributions of the target and the
fragmentation functions of the longitudinally polarized quark to
longitudinally polarized $\Lambda$. Thus it is possible to make a
systematic study of the quark helicity and transversity
distributions of nucleons, and of the polarized quark to polarized
$\Lambda$ fragmentations, by using the available facilities, such
as COMPASS, HERMES and SMC, on $\Lambda$ fragmentation in charged
lepton DIS on both longitudinally and transversely polarized
nucleon targets. The target nucleon can be chosen to be a proton
or a neutron (experimentally through $^2$H and $^3$He targets)
respectively, and this can add additional information for a clear
distinction of different predictions.

We now look at the quark to $\Lambda$ fragmentation functions
$D_q^{\Lambda}(z)$. Recently there has been progress in
understanding the quark to $\Lambda$ fragmentations \cite{MSSY} by
using the Gribov-Lipatov (GL) relation \cite{GLR}
\begin{equation}
D_q^h(z) \sim z \, q_h(z)~ \label{GLR}
\end{equation}
in order to connect the fragmentation functions with the
distribution functions. This relation, where $D_q^h(z)$ is the
fragmentation function for a quark $q$ splitting into a hadron $h$
with longitudinal momentum fraction $z$, and $q_h(z)$ is the quark
distribution of finding the quark $q$ inside the hadron $h$
carrying a momentum fraction $x=z$, is only known to be valid near
$z \to 1$ on an energy scale $Q^2_0$ in leading order
approximation \cite{BRV00}. However, predictions of $\Lambda$
polarizations \cite{MSSY} based on quark distributions of the
$\Lambda$ in the SU(6) quark-spectator-diquark model and in the
pQCD based counting rule analysis, have been found to be supported
by all available data from longitudinally polarized $\Lambda$
fragmentations in $e^+e^-$-annihilation
\cite{ALEPH96,DELPHI95,OPAL97}, polarized charged lepton DIS
process \cite{HERMES,E665}, and most recently, neutrino
(antineutrino) DIS process \cite{NOMAD}. Thus it is natural to
extend the same kind of analysis from longitudinally to
transversely polarized cases, and then check the validity of the
method by comparing theoretical predictions with experimental
data. Such an analysis can also serve as a theoretical guidance to
design future experiments.

The SU(6) quark-spectator-diquark model \cite{Ma96,Fey72,DQM}
starts from the three quark  SU(6) quark model wavefunction of the
baryon, and if anyone of the quarks is probed,
one reorganizes the
other two quarks in terms of two quark wavefunctions with spin 0
or 1 (scalar and vector diquarks), i.e., the diquark serves as an
effective particle, called the spectator. Some
non-perturbative effects such as gluon exchanges between the two
spectator quarks or other non-perturbative gluon effects in the
hadronic debris can be effectively taken into account by the mass
of the diquark spectator. The mass difference between the scalar
and vector diquarks has been shown to be important for producing
consistency with experimental observations of the ratio
$F_2^n(x)/F_2^p(x)=1/4$ at $x \to 1$ found in the early
experiments \cite{Fey72,DQM}, and also for the
proton and neutron polarized spin
dependent structure functions at
large $x$ \cite{Ma96,DQM}. The light-cone SU(6)
quark-spectator-diquark model \cite{Ma96} is an extended version of
this framework, taking into account the Melosh-Wigner
rotation effects \cite{Ma91b,Sch97}, in order to built up the quark
helicity and transversity distributions of the nucleon.
A detailed
discussion of quark helicity and transversity distributions in the
light-cone SU(6) quark-diquark model can be found in
Ref.~\cite{Ma98a}. It has been also shown recently \cite{MSY9}
that the predicted $x$-dependent transversity distributions are
compatible with the available HERMES data for the azimuthal
asymmetry \cite{HERMES00}. The application of the model for
discussing the quark helicity distributions of the $\Lambda$ can
be found in Refs.~\cite{MSSY}, where it is shown that the u and d
quarks inside the $\Lambda$ should be positively polarized at
large $x$, although their net spin contributions to the $\Lambda$
polarization might be zero or negative, and such a prediction was
found \cite{MSSY} to be in good agreement with the experimental
data. The extension of this framework to the quark transversity
distributions is straightforward, since one only needs to replace
the Melosh-Wigner rotation factor for helicity by that for
transversity \cite{Sch97,Ma98a}. We found similar qualitative
features between the helicity and transversity distributions for
each quark flavor, as can be seen from Figs.~\ref{mssy8f1} and
\ref{mssy8f2}, where the ratios $\Delta q(x)/q(x)$ and $\delta
q(x)/q(x)$ for the valence quarks of both proton
(Fig.~\ref{mssy8f1}) and $\Lambda$ (Fig.~\ref{mssy8f2}) are
presented.

\begin{figure}
\begin{center}
\leavevmode {\epsfysize=7cm \epsffile{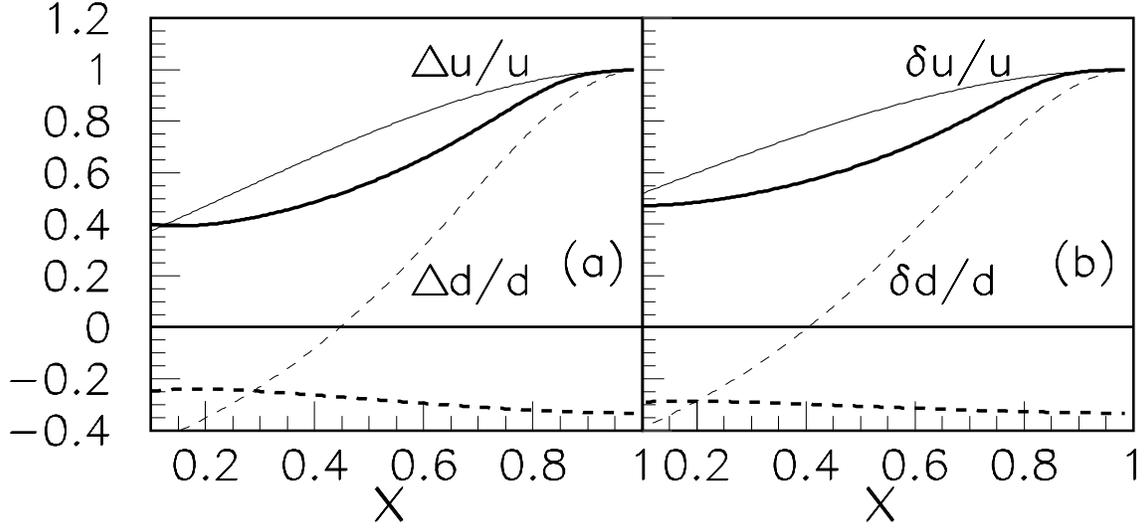}}
\end{center}
\caption[*]{\baselineskip 13pt The predicted ratios: (a) $\Delta
q(x)/q(x)$, and (b) $\delta q(x)/q(x)$, for proton in the
quark-diquark model (thick curves) and the pQCD based model
(thin curves). Solid curves are for $u$ valence quarks and
dashed curves are for $d$ valence quark.} \label{mssy8f1}
\end{figure}

\begin{figure}
\begin{center}
\leavevmode {\epsfysize=7cm \epsffile{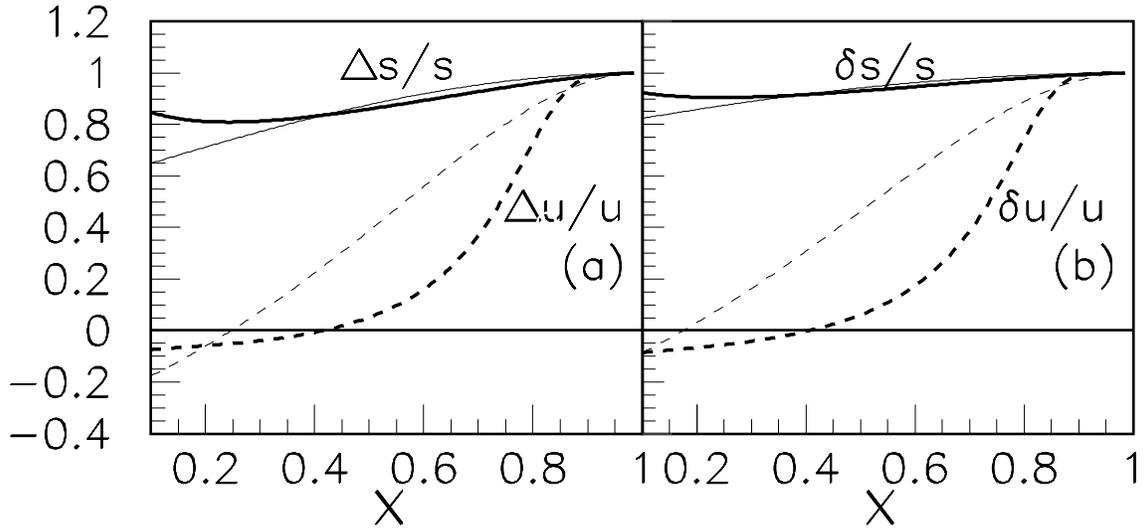}}
\end{center}
\caption[*]{\baselineskip 13pt The predicted ratios: (a) $\Delta
q(x)/q(x)$, and (b) $\delta q(x)/q(x)$, for $\Lambda$ in the
quark-diquark model (thick curves) and the pQCD based model
(thin curves). Solid curves are for $s$ valence quarks and
dashed curves are for $u$ and $d$ valence quarks.}
\label{mssy8f2}
\end{figure}

We notice that the $d$ quark in the proton is predicted to have a
negative quark helicity distribution at $x \to 1$, and this
feature is different from  the pQCD counting rule prediction of
``helicity retention", which means that the helicity of a valence
quark will match that of the parent hadron at large $x$.
Explicitly, the quark helicity distributions of a hadron $h$ have
been shown to satisfy the counting rule \cite{countingr},
\begin{equation}
q_h(x) \sim (1-x)^p, \label{pl}
\end{equation}
where
\begin{equation}
p=2 n-1 +2 \Delta S_z.
\end{equation}
Here $n$ is the minimal number of the spectator quarks, and
$\Delta S_z=|S_z^q-S_z^h|=0$ or $1$ for parallel or anti-parallel
quark and hadron helicities, respectively \cite{Bro95}. Therefore
the anti-parallel helicity quark distributions are suppressed by a
relative factor $(1-x)^2$, and consequently $\Delta q(x)/q(x) \to
1$ as $x \to 1$. Taking only the leading term, we can  write the
quark helicity distributions of the valence quarks as
\begin{equation}
\begin{array}{cllr}
&q^{\uparrow}_{i}(x)=\frac{\tilde{A}_{q_{i}}}{B_3}
x^{-\frac{1}{2}}(1-x)^3;\\
&q^{\downarrow}_{i}(x)=\frac{\tilde{C}_{q_{i}}}{B_5}
x^{-\frac{1}{2}}(1-x)^5,
\end{array}
\label{case2}
\end{equation}
where $\tilde{A}_{q}+\tilde{C}_{q}=N_q$ is the valence quark
number for quark $q$,  $B_n=B(1/2,n+1)$ is the $\beta$-function
defined by $B(1-\alpha,n+1)=\int_0^1 x^{-\alpha}(1-x)^{n} {\mathrm
d} x$ for $\alpha=1/2$, and $B_3=32/35$ and $B_5=512/693$. The
application of the pQCD counting rule analysis to discuss the
unpolarized and polarized structure functions of nucleons can be
found in Ref.~\cite{Bro95}, and the extension to the $\Lambda$ can
be found in Refs.~\cite{MSSY}. The $u$ and $d$ quarks inside the
$\Lambda$ are also predicted to be positively polarized at large
$x$ \cite{MSSY}, just as in the quark-diquark model prediction. It
is interesting that the predictions based on the pQCD based
counting rule analysis are also found \cite{MSSY} to be in
agreement with the experimental data, after some adjustment to the
parameters with higher order terms included.

The quark transversity distributions are closely related to the
quark helicity distributions. A useful inequality has been
obtained \cite{Soffer}, which constrains the quark transversity
distributions by the quark unpolarized and polarized
distributions, and there also exists an approximate relation
\cite{Ma98a} which connects the quark transversity distributions
with the quark helicity and spin distributions. Two sum rules
\cite{Ma98a}, connecting the integrated quark transversities with
some measured quantities and two model correction factors with
limited uncertainties, have been also recently obtained. For
example, if we assume the saturation of the inequality
\cite{Soffer}
\begin{equation}
2 |\delta q(x)| \le  q(x)+ \Delta q(x),
\label{SI}
\end{equation}
then we obtain $\delta q=\frac{1}{2}\left[q(x)+\Delta q(x)\right]
=q^{\uparrow}(x)$, and this suggests that in general we may
express $\delta q(x)$ in terms of $q^{\uparrow}(x)$ and
$q^{\downarrow}(x)$. All these considerations indicate that it is
convenient to parameterize the valence quark transversity
distributions in a similar form as the helicity distributions.
Therefore we use as a second model
\begin{equation}
\delta q(x)=\frac{\hat{A}_{q}}{B_3}
x^{-\frac{1}{2}}(1-x)^3-\frac{\hat{C}_{q}}{B_5}
x^{-\frac{1}{2}}(1-x)^5,
\end{equation}
which clearly satisfies the inequality (\ref{SI}). These quark
transversity distributions are constrained by $\delta Q=
\int_0^1\delta q(x) {\mathrm d} x $ from the two sum rules in
Ref.~\cite{Ma98a}. We also take $\hat{A}_{q}+\hat{C}_{q}=N_q$ as
in the case of the helicity distributions, in order to reduce the
number of uncertain parameters. Besides, all quark distributions
for the valence quarks of nucleons and the $\Lambda$ are assumed
to be connected between each other by the SU(3) symmetry relation
\begin{equation}
\begin{array}{lllc}
u^p=d^n=\frac{2}{3}u^{\Lambda}+\frac{4}{3}s^{\Lambda};\\
d^p=u^n=\frac{4}{3}u^{\Lambda}-\frac{1}{3}s^{\Lambda}.
\end{array}
\end{equation}
With the inputs of the quark helicity sum $\Sigma=\Delta U+ \Delta
D +\Delta S \approx 0.3$, the Bjorken sum rule
$\Gamma^p-\Gamma^n=\frac{1}{6}(\Delta U -\Delta D)=
\frac{1}{6}g_A/g_V \approx 0.2$, both obtained in charged lepton
DIS experiments \cite{Ma98a}, and taking the two model correction
factors both to be equal to 1 for the two sum rules of quark
transversities \cite{Ma98a}, we obtain $\Delta U=0.75$, $\Delta
D=-0.45$, $\delta U=1.04$, and $\delta D=-0.39$ for the proton,
assuming $\Delta S=0$. Such a scenario should be able to reflect
the bulk features of the valence quarks for the octet baryons,
although it might be too rough for their sea content. The $\delta
U$ and $\delta D$ so obtained are compatible with those from a
chiral soliton model \cite{CSM}. We may readjust the values when
experimental constraints become available, or if we believe other
models are more reasonable \cite{Ma98a}. It is encouraging that
the obtained transversity distributions for the nucleons have been
found to give consistent descriptions \cite{MSY9} of the available
HERMES data for the azimuthal asymmetry. The parameters for the
nucleons and $\Lambda$ quark distributions can be found in Table
1. The ratios $\Delta q(x)/q(x)$ and $\delta q(x)/q(x)$ for the
valence quarks of the proton and the $\Lambda$ in the pQCD based
model are also presented in Figs.~\ref{mssy8f1} and \ref{mssy8f2},
respectively. Notice that the helicity and transversity
distributions are close to each other at large $x$. This comes
from the fact that the Wigner-Melosh rotation factors reduce to 1
at the limit $x \to 1$.

\vspace{0.5cm}

\centerline{Table 1~~ The  parameters for quark distributions of
the nucleon and $\Lambda$
in the pQCD based model}

\vspace{0.3cm}

\begin{footnotesize}
\begin{center}
\begin{tabular}{|c||c|c||c|c|c|c||c|c|c|c|}\hline
 Baryon & $q_1$ & $q_2$ & $\tilde{A}_{q_1}$
 &$\tilde{C}_{q_1}$ &$\tilde{A}_{q_2}$
 &$\tilde{C}_{q_2}$ & $\hat{A}_{q_1}$
 &$\hat{C}_{q_1}$ &$\hat{A}_{q_2}$
 &$\hat{C}_{q_2}$ \\ \hline
 p & u & d & 1.375 & 0.625 & 0.275 & 0.725 &1.52&0.48&0.305&0.695\\
\hline n & d & u & 1.375 & 0.625 & 0.275 &
0.725&1.52&0.48&0.305&0.695
\\ \hline
$\Lambda$ & s & u(d) & 0.825 & 0.175 & 0.4125 & 0.5875
&0.912&0.088&0.457&0.543
\\ \hline
\end{tabular}
\end{center}
\end{footnotesize}
\vspace{0.5cm}

For $\Lambda$ production in the current fragmentation region along
the virtual photon direction, the spin transfer to the
longitudinal polarized $\Lambda$ is written as \cite{Jaf96,Lu95}
\begin{equation}
A^{\Lambda}(x,z)= \frac{\sum\limits_{q} e_q^2 \Delta q^N(x,Q^2)
\Delta D_q^\Lambda(z,Q^2) } {\sum\limits_{q} e_q^2 q^N (x,Q^2)
D^\Lambda_q(z,Q^2)}~ \label{LST}
\end{equation}
for charged lepton DIS on a longitudinally polarized nucleon $N$
target, and that to the transversely polarized $\Lambda$
is written as \cite{Art90,Jaf96}
\begin{equation}
\hat{A}^{\Lambda}(x,z)= \frac{\sum\limits_{q} e_q^2 \delta
q^N(x,Q^2) \delta D_q^\Lambda(z,Q^2) } {\sum\limits_{q} e_q^2 q^N
(x,Q^2) D^\Lambda_q(z,Q^2)}~ \label{TST}
\end{equation}
for charged lepton DIS on a transversely polarized nucleon $N$
target. Now we have the quark distributions $q(x)$, $\Delta q(x)$,
and $\delta q(x)$ for the valence quarks of nucleons and the
$\Lambda$ in both the SU(6) quark-diquark model and the pQCD
inspired analysis. For the quark to $\Lambda$ fragmentation
functions $D_q^{\Lambda}(z)$, $\Delta D_q^{\Lambda}(z)$, and
$\delta D_q^{\Lambda}(z)$, we use the Gribov-Lipatov relation
Eq.~(\ref{GLR}), in order to connect them with the corresponding
quark distributions of the $\Lambda$ in the two models. Therefore
we have the necessary inputs for a first numerical evaluation of
the two spin transfers  Eqs.~(\ref{LST}) and (\ref{TST}) in the
large $x$ and $z$ regions, where the valence quarks are dominant
inside the baryons. Extension to the small $x$ region requires the
knowledge of quark helicity and transversity distributions of the
target in this region, where we may use theoretical estimations or
parametrizations from other kinds of experiments as inputs.
Similarly, we can also use other experiments or theoretical
considerations to constrain the various quark to $\Lambda$
fragmentation functions, and extend our knowledge from the large
$z$ region to the small $z$ region. From the previous successful
predictions \cite{MSSY} of longitudinal $\Lambda$ polarizations,
supported by all available data, we expect that our results will
have some predictive power even in the small $z$ region.
Furthermore, by using the measured spin transfers for both
$A^{\Lambda}(x,z)$ and $\hat{A}^{\Lambda}(x,z)$, we can double
check our predictions from different models, and get a deeper
insight into the spin structure of both nucleons and the
$\Lambda$.

In the nucleon target, there are only $u$ and $d$ valence quarks,
therefore the dominant contribution to the two spin transfers
$A^{\Lambda}(x,z)$ and $\hat{A}^{\Lambda}(x,z)$ should come from
the $u$ and $d$ quark contributions in the large $x$ and $z$
regions. In the specific case of the proton target, the $u$ quarks
are dominant inside the target, its squared charges is 4/9, larger
than 1/4 of the $d$ quark, and also the ratios $\Delta u(x)/u(x)$
and $\delta u(x)/u(x)$ are positive values close to 1, which
causes the dominance of $u$ quark contributions inside the target.
Therefore the main features of the two spin transfers
$A^{\Lambda}(x,z)$ and $\hat{A}^{\Lambda}(x,z)$ are mainly
determined by the ratios $\Delta
D_u^{\Lambda}(z)/D_u^{\Lambda}(z)$  and $\delta
D_u^{\Lambda}(z)/D_u^{\Lambda}(z)$, as can be seen from Fig.~3.
Thus we can check the predicted $\Delta
D_u^{\Lambda}(z)/D_u^{\Lambda}(z)$ and $\delta
D_u^{\Lambda}(z)/D_u^{\Lambda}(z)$ by the measured spin transfers
$A^{\Lambda}(x,z)$  and $\hat{A}^{\Lambda}(x,z)$ from a proton
target.

\begin{figure}
\begin{center}
\leavevmode {\epsfysize=16cm \epsffile{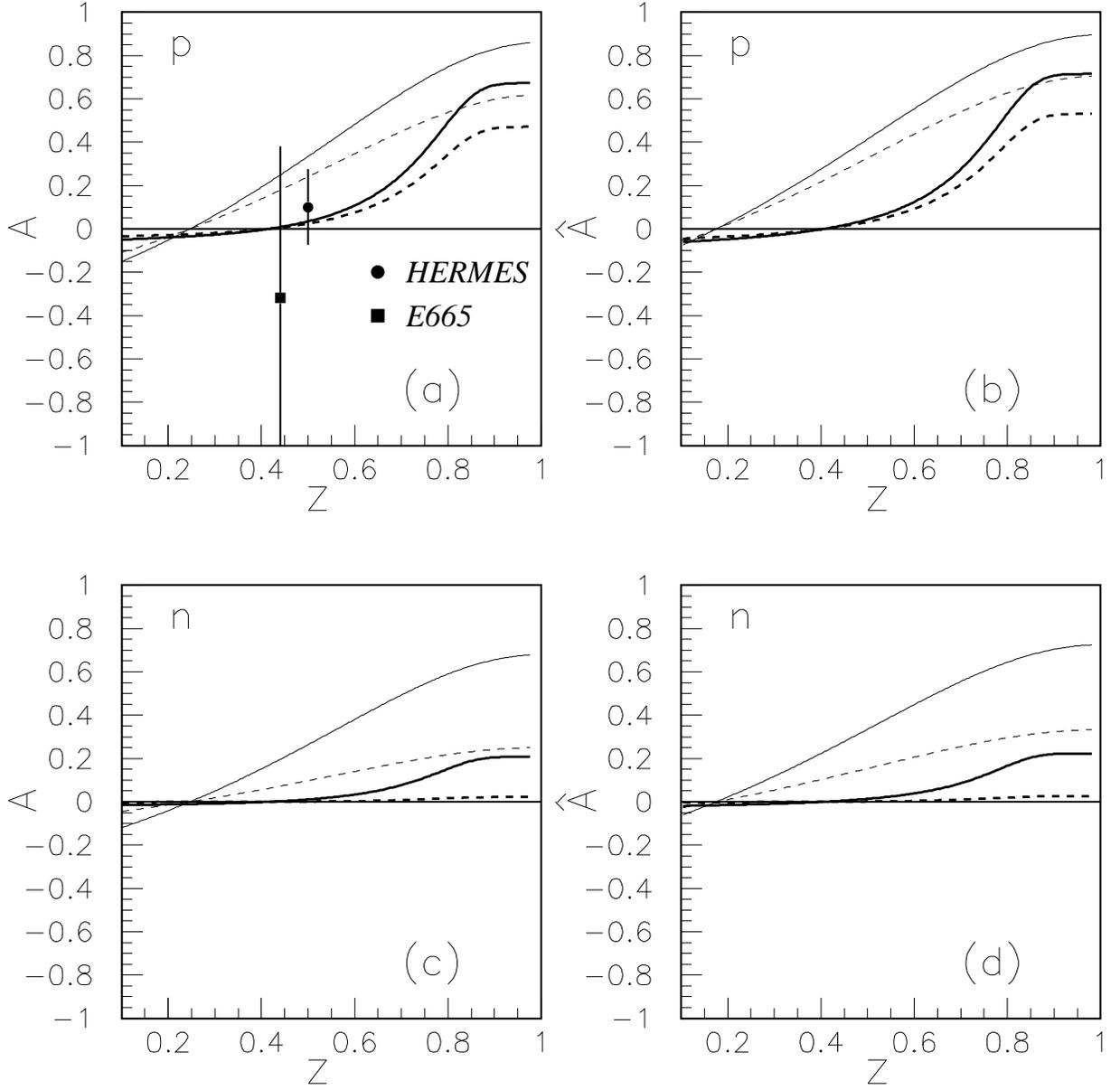}}
\end{center}
\caption[*]{\baselineskip 13pt The $x$-integrated spin transfers
$A^{\Lambda}(x,z)$  and $\hat{A}^{\Lambda}(x,z)$ of $\Lambda$
production in charged lepton DIS process on the longitudinally and
transversely polarized proton and neutron targets, with the
integrated $x$ range of $0.6 \to 1$ for the solid curves and $0.3
\to 1$ for the dashed curves. The thick curves correspond
to the results with quark distributions and fragmentation
functions from the quark-diquark model and the thin curves
correspond to these from the pQCD based model. The data are
taken by E665 \cite{E665} and HERMES \cite{HERMES} collaborations.
Notice that the cuts of the data are slightly different from
that of the prediction, but this does not change the qualitative
trends.
}\label{mssy8f3}
\end{figure}

\begin{figure}
\begin{center}
\leavevmode {\epsfysize=14cm \epsffile{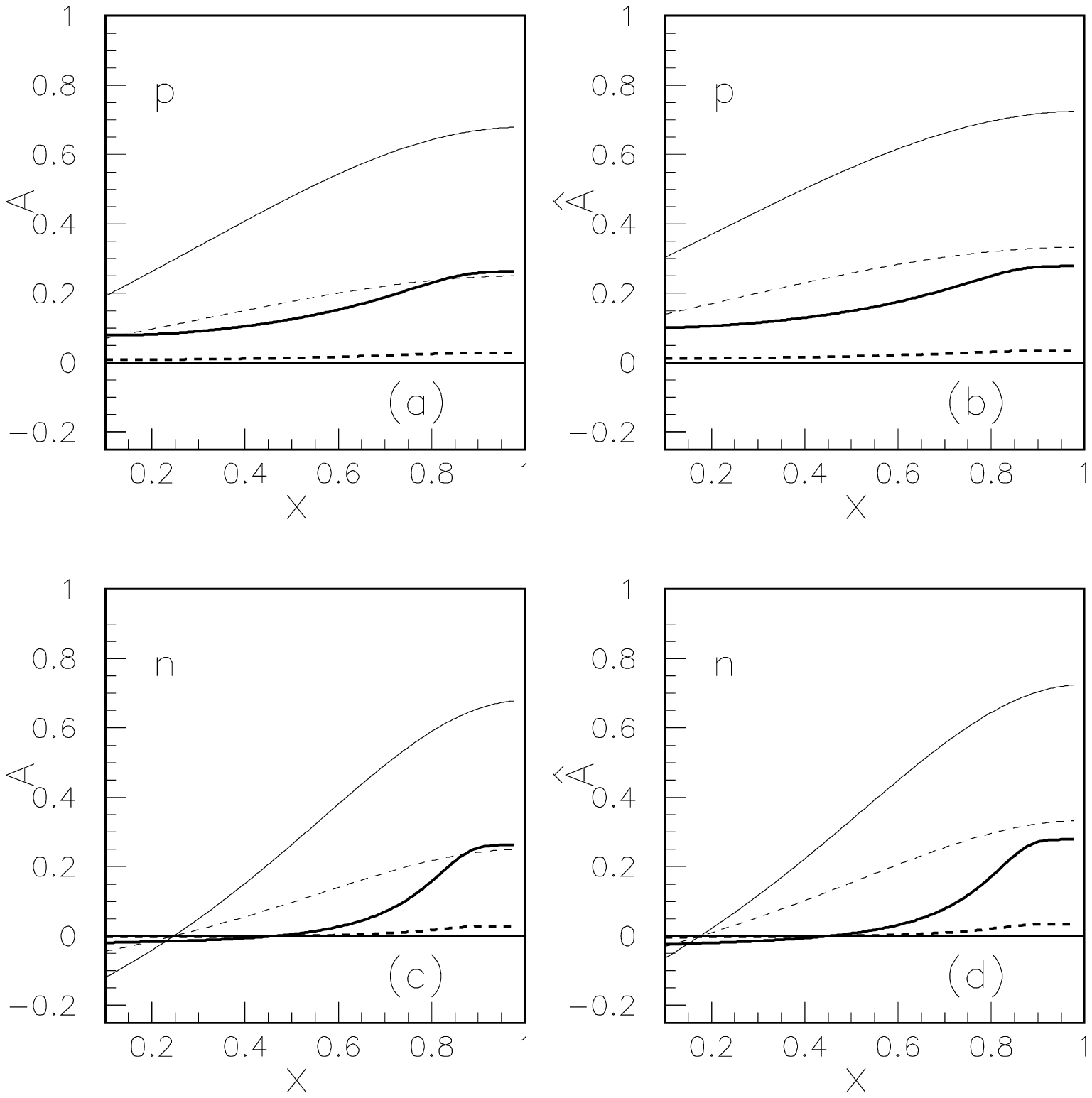}}
\end{center}
\caption[*]{\baselineskip 13pt The $z$-integrated spin transfers
$A^{\Lambda}(x,z)$  and $\hat{A}^{\Lambda}(x,z)$ of $\Lambda$
production in charged lepton DIS process on the longitudinally and
transversely polarized proton and neutron targets, with the
integrated $z$ range of $0.6 \to 1$ for the solid curves and $0.3
\to 1$ for the dashed curves. The thick curves correspond
to the results with quark distributions and fragmentation
functions from the quark-diquark model and the thin curves
correspond to these from the pQCD based model.
}\label{mssy8f4}
\end{figure}

The two models have qualitatively similar features for the ratios
$\Delta u(x)/u(x)$ and $\delta u(x)/u(x)$ for the $\Lambda$, and
consequently, we have qualitatively similar $\Delta
D_q^{\Lambda}(z)/D_q^{\Lambda}(z)$ and $\delta
D_q^{\Lambda}(z)/D_q^{\Lambda}(z)$ for the inputs to
Eqs.~(\ref{LST}) and (\ref{TST}), as can be seen from
Fig.~\ref{mssy8f2}.
In the non-relativistic model, the helicity and transversity distributions
are the same as the quark spin distributions in the quark model.
Therefore the difference between the helicity and transversity
distributions reflects the quark relativistic motion inside the nucleon.
As we mentioned before, the helicity and transversity
distributions are close to each other at large $x$, because the
Wigner-Melosh rotation factors reduces to 1 at the limit $x \to
1$. As a consequence we find no much difference between  $A$ and
$\hat A$. However, at small and medium $x$ when the sea quark
contribution cannot be neglected, the different chiral properties
between the helicity and transversity distributions will show up,
and their difference is ideal in order to study the chiral
properties of the nucleon.

Also we have $u \leftrightarrow d$ symmetry for the quark to
$\Lambda$ fragmentation functions. This implies that any big
qualitative difference of our predictions between the proton and
neutron targets are not mainly produced by the different inputs of
various quark to $\Lambda$ fragmentation functions in the two
models, but by the $u$ and $d$ difference in the quark helicity
and transversity distributions of the targets. Therefore the
different trends between the predictions of the spin transfers for
the proton and neutron targets, as can be seen in
Fig.~\ref{mssy8f3}, come mainly from the difference of the quark
helicity and transversity distributions for nucleons in the two
models. This can be easily understood because the weights of
squared charges are different for $u$ and $d$ quarks, and in the
neutron target the less dominant $u$ valence quark has more
weight, therefore $\Delta u^n(x)/u^n(x)\to -1/3$ (which is $\Delta
d^p(x)/d^p(x)$ from isospin symmetry) provides a bigger
contribution than for the proton target. This indicates that the
predicted spin transfers for the neutron target are more
suppressed in the quark-diquark model, whereas they are less
suppressed in the pQCD based model, as can be confirmed by
Fig.~\ref{mssy8f3}. Thus we conclude that the spin transfers
$A^{\Lambda}(x,z)$ and $\hat{A}^{\Lambda}(x,z)$ measured in both
large $x$ and large $z$ regions for the proton and neutron targets
can provide a check of the two different predictions of the quark
helicity and transversity distributions for the less dominant $d$
valence quark in the proton. They can also be used to test the
prediction of positively polarized $u$ and $d$ quarks inside the
$\Lambda$ at large $x$ for both models.

There have be available data of the spin transfer to the
longitudinal polarized $\Lambda$ in charged lepton proton DIS
scattering by E665 \cite{E665} and HERMES \cite{HERMES}
collaborations respectively, and we can compare the data
with our predictions as shown in Fig.~\ref{mssy8f3}~(a).
The precision of the data is still rough and the data
are compatible with both model predictions at medium
to large $z$ range.
High precision experiments are needed in order to
make clear distinction between different predictions and
we notice that the physics of the $\Lambda$ polarization
is strongly emphasize in the forthcoming COMPASS experiment \cite{COM}.
We also present the spin
transfers integrated over $z$ in Fig.~\ref{mssy8f4}, and find that
the $x$ dependence is not strong for the proton target,
especially for the quark-diquark
model in which the $x$ dependence of the ratios $\Delta q(x)/q(x)$
and $\delta q(x)/q(x)$ is not strong.
Therefore we can use a wide integrated $x$ range to increase
the statistics of the data.
We should stress that our
predictions should be considered to be valid more qualitatively
than quantitatively, especially for the pQCD based model. In this
case there is still freedom to include higher order terms and to
adjust the parameters of the pQCD based model from the constraints
of the data. Varying $x$ and $z$ in different regions can provide
us more information concerning the quark helicity and transversity
distributions of the target, as well as the quark to $\Lambda$
fragmentation functions.
We would like to mention that
similar analysis can be also made for
the spin transfers of other members of the octet baryons.
The analysis
and main conclusion for the spin transfers of the octet
baryons fragmentation, when the target nucleon is polarized either
longitudinally or transversely, should be similar to those found
in hadron longitudinal polarizations of the octet baryons in
polarized charged lepton DIS processes
\cite{MSSY}.

In conclusion, we showed in this paper that the $\Lambda$-hyperon
fragmentation in charged lepton DIS on the polarized nucleon
target can provide sensitive information concerning the quark
helicity and transversity distributions for both nucleons and the
$\Lambda$-hyperon at large $x$. We calculated the spin transfers
of the produced $\Lambda$ when the target nucleon is polarized
either longitudinally or transversely, with the nucleon and
$\Lambda$ quark distributions evaluated both in the SU(6)
quark-spectator-diquark model and in a pQCD based model. We found
that the predicted spin transfers have quite different behaviors
for the proton and neutron targets in the two models, and this can
provide a sensitive test of different predictions for the quark
helicity and transversity distributions for the $d$ valence quark
of the proton at large $x$.

{\bf Acknowledgments: } This work is partially supported by
National Natural Science Foundation of China under Grant Numbers
10025523,
19975052, and 19875024, by Fondecyt (Chile) postdoctoral fellowship
3990048, by the cooperation programmes Ecos-Conicyt and CNRS-
Conicyt between France and Chile, by Fondecyt (Chile) grants
1990806 and 8000017, and by a C\'atedra Presidencial (Chile).

\newpage
\noindent
{\bf Erratum:}

\noindent Published in
Phys. Rev. {\bf D 64}, 099901 (2001)

\noindent
{\bf Helicity and transversity distributions of the nucleon and $\Lambda$
hyperon from $\Lambda$ fragmentation}

\noindent
[Phys. Rev. D 64, 014017 (2001)]

\noindent
Bo-Qiang Ma, Ivan Schmidt, Jacques Soffer, and Jian-Jun Yang

\noindent
PACS number(s): 14.20.-c, 13.85.Ni, 13.87.Fh, 13.88.+e

The HERMES and E665 data in Fig.3 and the related discussions should
be removed. The reason is that the spin transfer discussed in the paper
is for {\it unpolarized} charged lepton DIS process on a 
{\it longitudinally polarized} target,
whereas the HERMES and E665 data are for {\it polarized} lepton DIS process on an
{\it unpolarized} target. Therefore the theoretical predictions presented 
in this paper have not been measured yet. 
The predictions and conclusions of the paper remain unchanged.

\end{document}